\documentclass[aps,twocolumn, showpacs,amssymb]{revtex4}
\newcommand{\be}{\begin{equation}}
\newcommand{\ee}{\end{equation}}

\makeatletter
\def\@pnumwidth{2em}
\makeatother
\usepackage{graphicx}
\usepackage{dcolumn}
\usepackage{bm}

\begin{document}
\title{Power law eigenvalue density, scaling and critical random matrix ensembles}
\author{K. A. Muttalib}
\email{muttalib@phys.ufl.edu}
\affiliation{Department of Physics, University of
Florida, P.O. Box 118440, Gainesville, FL 32611-8440}
\author{Mourad E.H. Ismail}
\email{ismail@math.ucf.edu}
\affiliation{Department of Mathematics, University of Central
Florida, Orlando, FL 32816}

\begin{abstract}
We consider a class of rotationally invariant unitary random
matrix ensembles where the eigenvalue density falls off as an
inverse power law. Under a new scaling appropriate for such power
law densities (different from the scaling required in Gaussian
random matrix ensembles), we calculate exactly the two-level
kernel that determines all eigenvalue correlations. We show that
such ensembles belong to the class of critical ensembles.

\end{abstract}


\maketitle

\section{Introduction}

Gaussian random matrix ensembles, introduced by Wigner and Dyson
\cite{wigner}, have been studied extensively over half a century
in the context of nuclear physics, atomic and molecular physics,
condensed matter physics as well as particle physics
\cite{review}. The wide applicability results from the universal
properties of the correlation functions of the eigenvalues, once
they are appropriately scaled. Thus, e.g., the correlation
functions of the eigenvalues $x$ of an ensemble of $N\times N$
hermitian matrices become, after proper scaling, independent of
the size of the matrix or of the details of the microscopic
distribution from which the matrix elements are drawn. For a
Gaussian distribution, the density is a semicircle given by
$\sigma(x)=\sqrt{2N-x^2}/\pi$; this requires, for universality,
the double scaling limit \be N\rightarrow\infty, \;\;\; x
\rightarrow 0,\;\;\; u\equiv x\sqrt{N}\;\;\; \textrm{finite},
\label{scaling}\ee such that the density
$\bar{\sigma}(u)=\rm{constant}$ near the origin. This particular
double scaling is required for all Gaussian random matrix
ensembles of different symmetry classes \cite{mehta} in order to
have a universal large $N$ limit. In particular, for Gaussian
unitary ensembles, after a second trivial scaling
$\zeta=\sqrt{2}u/\pi$ such that $\tilde{\sigma}(\zeta)=1$, it
leads to the universal two-level kernel  \be
K^G(\zeta,\eta)=\frac{\sin[\pi(\zeta-\eta)]}{\pi(\zeta-\eta)}\ee
independent of $N$, which gives rise to universal eigenvalue
correlations like the nearest-neighbor spacing distribution or the
number variance, commonly known as the Wigner distributions.

Note that in the above double scaling limit, one is always
restricted to the eigenvalues far from the tails of the density. This is a highly
non-trivial feature of Gaussian random matrix ensembles. Indeed,
the eigenvalue correlations for Gaussian ensembles are deeply
related to the properties of Hermite polynomials \cite{mehta}, and
the above scaling is dictated by the asymptotic properties of
Hermite polynomials of order $N$ and argument $x$ in the large $N$
limit for small $x$. Since Hermite polynomials have a different
asymptotic behavior for large $x$ in the large $N$ limit
\cite{szego}, a different scaling is dictated for the Gaussian
ensembles near the edge of the semi-circle spectrum. Here
universality is recovered after a shift $x'=x-\sqrt{2N+1}$, and a
different double scaling limit near the new origin, namely
$N\rightarrow\infty$, $x'\rightarrow 0$, with $u \equiv x'N^{1/6}$
remaining finite \cite{forrester}. In this case one gets the so
called Airy kernel and the density decreases exponentially for
large $u$.

In a sense, the universality of the Gaussian random matrix
ensembles is a result of an underlying central limit theorem,
generalized to matrices \cite{CB}. Indeed, the universality
requires that the eigenvalues are confined `strongly enough' such
that the fluctuations remain bounded. For example, if the
probability density $P(H)$ for an $N\times N$ random Hamiltonian
$H$ is given by \be P(H) \propto \exp[-\beta Tr V(H)],\ee then in
a more general situation where the confining potential $V(H)$ is
not a Gaussian, the density may not be the same near the origin.
Even then, one can `unfold' the spectrum by choosing a variable in
which the density is unity, and as long as the confining potential
is `strong enough', the double scaling given by
Eq.~(1) always leads to the same two-level kernel and therefore
the same universal correlations. This is because for any such
rotationally invariant ensemble described by a confining
potential, the two-level kernel can be described in terms of
polynomials  orthogonal  with respect to a Freud-type weight
function $e^{-V(x)}$, with polynomial $V$,
(just as the Hermite polynomials arise for Gaussian
ensembles where the potential is $V(x)=x^2$)\cite{mehta}. Again,
the eigenvalue correlations are deeply related to the properties
of those orthogonal polynomials. However, all Freud-type
orthogonal polynomials have qualitatively similar asymptotic
behavior in the large $N$ limit. Therefore the same double scaling
limit leads to universality for all such different confining
potentials. Conversely, an arbitrary scaling (with e.g. an
arbitrary power law for $N$) does not lead to a universal
$N$-independent kernel for rotationally invariant random matrix
ensembles associated with the classical orthogonal polynomials.

On the other hand, it is known that the universality does break
down if the confinement potential $V(H)$ grows extremely slowly
with $H$ \cite{muttalib}, namely \be V(H) = \frac{1}{\gamma}\ln^2
H, \;\;\; H \gg 1. \ee This leads to the so called `critical'
ensembles \cite{kravtsov}, characterized by a one-parameter
generalization of the Gaussian two-level kernel, given by
\cite{muttalib,shapiro} \be
K^C(\zeta,\eta;\gamma)=\frac{\gamma}{2\pi}
\frac{\sin[\pi(\zeta-\eta)]}{\sinh[(\zeta-\eta)\gamma/2]}, \;\;\;
\gamma \ll 2\pi^2 .\ee The eigenvalue correlations for such
systems for various values of the parameter $\gamma$ have been
studied in the context of the Anderson transition in disordered
quantum conductors \cite{nishigaki}. The density of the
eigenvalues with the weak confinement potential (4) turns out to
be a constant independent of $N$ in the $N\rightarrow\infty$
limit, and the scaled variables are simply $\zeta=x/\gamma$.  Note
that the critical ensembles tend to the Gaussian ensembles in the
limit $\gamma\rightarrow 0$, in which case one is again restricted
to the correlations of the eigenvalues near the origin. However,
the scaling does not involve $N$. This difference in the scaling
behavior is the reason why critical ensembles could not be studied
using the classical orthogonal polynomials.

A second case where a different scaling beyond the scope of the
Freud-type  orthogonal polynomials is required is when the density
of the eigenvalues $x$ falls off as a power law for large $x$.
This indeed happens for a large class of random matrix ensembles
of recent interest, relevant for e.g. finance or scale free
systems that requires considering power law distributions with fat
tails \cite{laloux,burda}. In a pioneering study, Cizeau and
Bouchaud (CB) \cite{CB} introduced the so called `L\'evy matrix
ensembles' where the matrix elements are drawn from a power law
distribution according to $P(H_{ij})$ with \be P(H_{ij})\sim
\frac{H^{\mu}_0}{|H{ij}|^{1+\mu}}, \;\;\; H_{ij} \gg 1.\ee Here
$H_0$ is of order $H_{ij}$ and the parameter $\mu \ge 0$. The
eigenvalue density in this case was shown to fall off with the
same exponent, as $1/x^{1+\mu}$. In particular for $0< \mu <1$,
the distribution has \textit{diverging} variance and hence the
underlying central limit theorem must now be modified according to
the theorems of L\'evy and Gnedenko \cite{bouchaud}. Indeed,
several numerical work have shown \cite{bouchaud,choi} that the
universality of the Gaussian ensembles does break down when the
density of eigenvalues follows a power law with exponent less than
2, the signature of which is apparent in the numerically obtained
nearest-neighbor spacing distribution and the number variance. The
eigenvalue densities for certain L\'evy ensembles have been
obtained analytically \cite{burda}, but so far it has not been
possible to evaluate the two-level kernel which gives the two or
higher level eigenvalue correlations. Since the universality
requires `unfolding' where the density is made uniform, it is
obviously of more interest to obtain the two-level kernel for a
L\'evy-like ensemble, specially where the microscopic distribution
leads to a power law eigenvalue-density which falls off with an
exponent less than 2.

Clearly, if the density of eigenvalues falls off with a power law
and we are interested in the properties of the large eigenvalues
in the tails, any possible universality of the associated random
matrix ensembles cannot be expected in the same double scaling
limit used for the Gaussian ensembles, either at the origin or
near the semi-circle edge. In [\onlinecite{CB}], CB proposed a
double scaling limit \be N\rightarrow\infty, \;\;\;
x\rightarrow\infty, \;\;\; u\equiv x/N^{\alpha} \;\;\;
\textrm{finite},\ee where $\alpha>0$ is related to the power law
exponent, in which the L\'evy ensembles might become universal. We
will call this the CB scaling. As mentioned above, this scaling
cannot be obtained naturally within the scheme of classical or
Freud-type orthogonal polynomials. This means that the most
powerful approach to study eigenvalue correlations of random
matrix ensembles, namely the method of orthogonal polynomials
\cite{mehta}, seem to be inapplicable for L\'evy-type random
matrices. It is for this reason that progress has been slow in
detailed analytic studies of eigenvalue correlations of  random
matrices with power law densities.

On the other hand, although in most studies it has been implicitly
assumed that a power law density of eigenvalues requires choosing
matrix elements from a power law distribution of the type (6), we
find that at least for the inverse power law with exponent 1, this
is not the case. Indeed, we find that a weakly confined
log-squared potential that gives rise to the critical ensemble
with a constant density for an $N$-independent scaling also gives
rise to a inverse power law density under the CB scaling. It turns
out that the weakly confined log-squared potential is exactly
solvable even in the new CB scaling regime and the two-level
kernel in this scaling regime has a well-defined limit as
$N\rightarrow\infty$.

\section{The model}

In the present work we use the solvability of the log squared
potential to obtain analytically, exactly and explicitly, the
two-level kernel for a random matrix ensemble with density
$\sigma(x) \propto 1/x$ in the CB double scaling limit given in
Eq.~(7) where $\alpha$ is an arbitrary positive parameter.
Specifically, we use the model characterized by the confinement
potential \be\label{V} V(x|q)=\frac{2}{|\ln
q|}[\ln(x+\sqrt{1+x^2}]^2\ee where $q$ is a parameter, $0 <q \le 1
$. The joint probability distribution of the N eigenvalues can
then be written as \cite{mehta} \be P(\{x\})=C_N(q)\prod_{i<j}^N
(x_i-x_j)^2\prod_{i=1}^N e^{-V(x|q)}\ee where $C_N(q)$ is a
constant and the product $\prod_{i<j}^N (x_i-x_j)^2$ is the
standard Vandermonde factor for unitary ensembles. The orthogonal
polynomials corresponding to the weight function \be w_H(x|q)=
q^{1/8}\sqrt{\frac{-2}{\pi\ln q}}e^{-V(x|q)}\ee for the above
confining potential, in terms of which the two-level kernel can be
written down exactly, are the Ismail--Masson $q^{-1}$-Hermite
polynomials \cite{ismail} first considered in the context of
random matrices in [\onlinecite{muttalib}], where only an
N-independent scaling was considered. The reason for the choice of
the potential (\ref{V}) is that a very general asymptotic relation
for the corresponding Ismail--Masson polynomials has been obtained
recently \cite{ismail-zhang}, such that the new CB scaling can be
implemented explicitly without compromising the solvability of the
model. In particular, for \be x_n(t,u) = \frac{1}{2}(q^{-nt}u -
q^{nt}/u) \ee and $0<t<1/2$ \cite{note} it has been shown that the
normalized Ismail--Masson polynomials $\tilde{h}_n(x_n|q)$ has the
large $n$ limit given by
\begin{eqnarray}
\nonumber \sqrt{w_H(x_n)}\;\tilde{h}_n(x_n) =
\frac{\sqrt{w_H(\sinh \ln u|q)}u^{2a}}{(-1)^{\lfloor m/2\rfloor}
(q;q)_{\infty}\sqrt{(q;q)_n}}\times \\*
q^{n/4}q^{a^2}\Theta(-u^2q^{2a},q)
\end{eqnarray}
Here $(q;q)_n=\prod_0^{\infty}(1-q^n)$, $m$ is an integer
 and $0\le\lambda<1$ has been
defined by the relation $m+\lambda=n(1-2t)$. We have also defined
$2a = \chi(m)+\lambda$ where $\chi(m)$ is 0(1) for $m=$even(odd).
The function \be
\Theta(z,q)=\sum_{n=-\infty}^{\infty}q^{n^2}z^n\ee is defined for
$z\neq 0$ for any complex $z$.

Let us define \be q\equiv e^{-\gamma}, \;\;\;\gamma \ge 0.\ee Note
that for $N\rightarrow\infty$ and large $u$, we can choose the
parameters $\gamma$ and $t$ to satisfy the CB scaling for
arbitrary $\alpha$. The choice  \be \gamma Nt=\alpha \ln N, \ee
when used in Eq.~(11), gives precisely the CB scaling $
u=x/N^{\alpha}$. Note that while for finite $0< t<1/2$ the above
condition requires $\gamma =2\alpha\ln (N)/N\rightarrow 0$ or
equivalently $q\rightarrow 1$, we are free to choose $t\sim \ln
(N)/N$ such that the scaling relation can be satisfied for finite
$\gamma$. In the following, we will outline the derivation of the
two-level kernel for $\gamma \ll \pi^2$.

\section{The two-level kernel}

The two-level Ismail--Masson kernel defined in terms of the orthonormal
polynomials $\tilde{h}_n(x|q)$ is given by
$K_N(x,y)=\sqrt{w_H(x_N)w_H(y_N)}
\sum_{k_=0}^{N-1}\tilde{h}_k(x|q)\tilde{h}_k(y|q)$. For large N it
can be written, using Christoffel-Darbeau formula, as
\begin{eqnarray}
\nonumber
K_N(x,y)=\sqrt{w_H(x_N)w_H(y_N)}\frac{k_N}{k_{N-1}}\times \\*
\frac{\tilde{h}_N(x_N|q)\tilde{h}_{N-1}(y_N|q)-
\tilde{h}_N(y_N|q)\tilde{h}_{N-1}(x_N|q)}{x-y}.
\end{eqnarray}
where $k_N$ is the coefficient of the term $x^N$ in the polynomial
of order $N$. In Eq.~(16) we can replace $N$ by $N-1$ to obtain
the asymptotic expression for the normalized polynomials
$\tilde{h}_{N-1}(x_{N-1}|q)$. However, the Christoffel-Darbeaux
formula involves $\tilde{h}_{N-1}(x_{N}|q)$. Similarly, we need
expression for the weight factor $w_H(x_{N-1}|q)$ in terms of
$w_H(x_{N}|q)$. We obtain these results by exploiting the scaling
relation \be x_{N-1}(t,q^{-t}u)=x_N(t,u),\ee which follows from
Eq.~(11). This gives, in terms of the scaled variables
$v=y/N^{\alpha}$,
\begin{eqnarray}
\nonumber \sqrt{w_H(y_N)}\;\tilde{h}_{N-1}(y_N) =
\frac{\sqrt{w_H(\sinh \ln (q^{-t} v)|q)}}{(-1)^{\lfloor
m'/2\rfloor} (q;q)_{\infty}\sqrt{(q;q)_{N-1}}}&\times& \\*
(q^{-t}v)^{2b+2t}q^{(N-1)/4}q^{(b+t)^2}\Theta(-v^2q^{2b},q)&\;&
\end{eqnarray}
where $m'+\lambda'=(N-1)(1-2t)$ and we have defined $2b =
\chi(m')+\lambda'-2t$.

The functions $\Theta(z,q)$ appearing in the above equations
cannot be evaluated easily. However, for $q$ near 1 (or $\gamma
\ll \pi^2$) which is the region we would be interested in (this
allows us to go to the Gaussian limit $q\rightarrow 1$), we can
use the imaginary transformation \cite{watson} \be \Theta(w,q) =
\sqrt{\frac{\pi}{\gamma}}
e^{\frac{\ln^2w}{4\gamma}}\Theta(e^{\frac{\pi\ln w}{i\gamma}},p);
\;\;\; p=e^{-\frac{\pi^2}{\gamma}} .\ee In this case the series in
powers of $p =e^{-\pi^2/\gamma}$ converges rapidly for $\gamma \ll
\pi^2$ and can be approximated quite well by keeping only the
dominant term. The transformation allows us to rewrite the kernel
as
\begin{eqnarray} \nonumber (x-y)\;K(x,y)=A
(q)e^{\frac{i\pi}{\gamma}\ln(uv)}
\sum_{k_1,k_2}p^{k^2_1+k^2_2-k_1-k_2}&\times& \\*
(pz_1)^{k_1}(pz_2)^{k_2}
[1-e^{-i2\pi(a-b)(k_1-k_2)}]&\;&\end{eqnarray} where
$pz_1=e^{i2\pi[a-\frac{1}{\gamma}\ln u]}$,
$pz_2=e^{i2\pi[b-\frac{1}{\gamma}\ln v]}$ and $A(q)$ is a constant
independent of $u,v$. We now make our first approximation, namely
that we keep only the two terms $k_1=0, k_2=1$ and $k_1=1, k_2=0$
in the double sum in Eq.~(20). This neglects terms of order $p^2$
or higher powers of $p$. Using the fact that the product $
(-1)^{\lfloor m/2\rfloor +\lfloor m'/2\rfloor}\cdot \sin[\pi(a-b)]
= 1 $ for all $a,b$, we finally obtain \be K(x,y) = K_0
q^{nt}\frac{\sin[\frac{\pi}{\gamma}\ln\frac{u}{v}]}{u-v}\ee where
\be K_0=\sqrt{\frac{2}{\pi\gamma}}\frac{2\pi}{\gamma}q^{-1/8}
\frac{e^{-\frac{\pi^2}{2\gamma}}}{(q;q)^3_{\infty}}.\ee In order
to estimate $K_0$, we use the Euler identity \cite{gasper} $
(q;q)_{\infty}=\prod_{n=1}^{\infty}(1-q^n)
=\sum_{n=-\infty}^{\infty} (-1)^nq^{n(3n+1)/2}$. The sum can be
evaluated using the Poisson summation formula to yield \be
(q;q)_{\infty}=\sqrt{\frac{2\pi}{3\gamma}}e^{\gamma/24}
\sum_{k=-\infty}^{\infty}e^{-\frac{\pi^2}{6\gamma}(2k+1)^2
-\frac{i\pi}{6}(2k+1)}.\ee Again we use the fact that $\gamma \ll
\pi^2$ and approximate the series by keeping only the $k=0$ and
$k=-1$ terms. This gives $(q;q)_{\infty}\approx
\sqrt{\frac{2\pi}{\gamma}}q^{-1/24}e^{-\frac{\pi^2}{6\gamma}}$.
This then leads to $K_0=1/\pi$.

\section{Results and discussion}

In terms of the scaled variables $u=x/N^{\alpha}$,
$v=y/N^{\alpha}$, the two-level kernel in the large $N$ limit is
given by \be K^P(u,v;\gamma) = \frac{1}{\pi}
\frac{\sin[\frac{\pi}{\gamma}\ln\frac{u}{v}]}{u-v}, \;\;\; 0<
\gamma \ll \pi^2\ee with density \be \sigma (u)\equiv K^P(u,u) =
\frac{1}{\gamma u}.\ee Thus the kernel is not universal, and
depends on one parameter that cannot be scaled out. Surprisingly,
once unfolded by changing variables to \be \zeta =\frac{1}{\gamma}
\ln u, \;\;\; \eta = \frac{1}{\gamma} \ln v \ee such that the
density becomes unity, $ \bar{\sigma} (\zeta)= 1,$ the scaled
kernel becomes identical to the critical kernel of Eq.~(5), \be
\bar{K}^P(\zeta,\eta;\gamma)\equiv \frac{K^P(\zeta,\eta;\gamma)}
{\sqrt{K^P(\zeta,\zeta;\gamma)K^P(\eta,\eta;\gamma)}}
=K^C(\zeta,\eta;\gamma).\ee Thus it shows that rotationally
invariant random matrix ensembles with inverse power law density
$\sigma (x)\propto 1/x$ belong to the class of critical ensembles,
albeit after a logarithmic variable transformation.

It is important to emphasize that although the current model
belongs to the same class of critical ensembles, it differs from
the $N$-independent scaling model in two important ways. First,
the $N$-dependent scaling introduces non-trivial $N$-dependent
logarithmic potential in addition to the log-squared potential of
the $N$-independent scaling model. Second, the logarithmic
variable transformation changes the interaction between the
eigenvalues. It is therefore not obvious that the present model
should lead to a critical ensemble. In the variable where both
models have weakly confining log-squared potential, the kernels
for the CB scaling and the $N$-independent scaling are given by
Eqs (24) and (5), respectively.

It has been conjectured before \cite{bertuola} that an inverse
power law density in a rotationally invariant random matrix
ensemble approaches the same behavior as that for a weakly
confined log-squared potential considered here. In this work we
show explicitly by calculating exactly the two-level kernel that
the inverse power case is a critical ensemble under the CB
scaling. However, it differs from the case of the weakly confined
log-squared potential with $N$-independent scaling in important
ways; in the variable where the present kernel becomes identical
to that for $N$-independent scaling, the confining potential is in
fact Gaussian.

The Ismail--Masson polynomials go back to the classical orthogonal
polynomials in the limit $q\rightarrow 1$. One can therefore hope
to implement the scaling for the Ismail--Masson polynomials and
then take the $q\rightarrow 1$ limit to obtain the corresponding
scaling for the Gaussian ensembles. It turns out that the inverse
power density $\sigma(u)\propto 1/u$ is very special and exists
only for $q < 1$. The unfolded kernel with constant density of
course has a well defined $q\rightarrow 1$ limit which is just the
sine kernel of the Gaussian ensemble. It should be possible to
obtain the two-level kernel for arbitrary density power law with
CB type scaling within our scheme, but it is not clear if a
well-defined $N\rightarrow\infty$ limit exists for the resulting
kernel. Here we presented only the inverse power density case,
which we show to be critical.

----------------------


\begin{thebibliography}{40}
\bibitem{note} The asymptotic relation for the Ismail--Masson polynomials
has been obtained in \cite{ismail-zhang} for all $t>0$. Here we use only
the result for $t<1/2$.
\end{thebibliography}
\end{document}